\documentclass{aastex}
\usepackage{spr-astr-addons}
\usepackage{url}

\RequirePackage{color}

\newcommand{\emaila}{wj@bao.ac.cn}

\begin{document}

\title{18-Months Operation of Lunar-based Ultraviolet Telescope: A Highly Stable Photometric Performance}
\slugcomment{Not to appear in Nonlearned J., 45.}
%% Running heads
\shorttitle{Highly Stable Photometric Performance of LUT}
\shortauthors{Wang et al.}

\author{J. Wang\altaffilmark{1,2}, X. M. Meng\altaffilmark{1,2}, X. H. Han\altaffilmark{1,2}, H. B. Cai\altaffilmark{1,2}, L. Cao\altaffilmark{1,2}, 
J. S. Deng\altaffilmark{1,2}, Y. L. Qiu\altaffilmark{1,2}, S. Wang\altaffilmark{1,2}, J. Y. Wei\altaffilmark{1,2}, J. Y. Hu\altaffilmark{1}}
%\affil{Affilation of First and Second Authors}
%\affil{Affilation of Third Author}
\email{\emaila}

\altaffiltext{1}{National Astronomical Observatories, Chinese Academy of Sciences}
\altaffiltext{2}{Key Laboratory of Space Astronomy and Technology, National Astronomical Observatories,
Chinese Academy of Sciences}
%\altaffiltext{3}{Third Alternate Affilation.}

\begin{abstract}

We here report the photometric performance of Lunar-based Ultraviolet telescope (LUT), the first 
robotic telescope working on the Moon, for its 18-months operation. In total, 17 IUE standards have been observed in 51 runs until
June 2015, which returns a highly stable photometric performance during the past 18 months (i.e., no evolution of 
photometric performance with time).
The magnitude zero point is determined to be $17.53\pm0.05$ mag,
which is not only highly consistent with the results based on its first 6-months operation, 
but also independent on the spectral type of the standard from which the 
magnitude zero point is determined. The implications of this
stable performance is discussed, and is useful for next generation lunar-based astronomical observations.

\end{abstract}


%\section*{}
%\label{sec:intro}

\section{Introduction}

As the first robotic astronomical telescope
working on the lunar surface in the history of mankind, Lunar-based Ultraviolet Telescope (LUT)
on board the Chinese first lunar lander
(Chang'E-3, Ip et al. 2014) has smoothly worked for 18 months on the Moon up to the time when this paper is
prepared. About 10,000 images have been acquired per month by LUT in the past 18 months.

A detailed description on the scientific goals and mission conception of LUT can be found in Cao et al. (2011).
Briefly speaking, the telescope is dedicated to 1) continuously monitor bright variable stars in the near-ultraviolet (NUV) band for as long as
a dozen days; 2) perform a sky survey at low Galactic latitude in the NUV band. 
LUT consists of a 2-dimensional gimbal with a flat mirror used for pointing a given target and a telescope with an aperture of 150mm. 
Both components are horizontally mounted in a cabin of the lander (see Figure 1 in Wang et al. 2015a for a cross sectional view). 
The telescope adopts a Ritchey-Chretien system with a focal ratio of 3.75, and is equipped with
an UV-enhanced 1024$\times$1024 AIMO CCD E2V47-20 as the detector.
The detector is operated in frame transfer mode, and can be thermoelectrically cooled by as much
as 40\symbol{23}C below its ambient temperature, which results in a typical operating temperature in a range 
from -40 (at dawn or dusk) to -20\symbol{23}C (at noon). 
The pixel size of the CCD is 13$\mu$m, which corresponds a pixel scale of
4.76\symbol{125}$\mathrm{pixel^{-1}}$  and results in a
filed-of-view of 1.36$\times$1.36$\mathrm{degrees^2}$. 
Two LED lamps with a center wavelength of
286 nm are equipped to provide an internal flat field. 
The efficiency curve of LUT that is measured in 
laboratory peaks at around 2500\AA\ with
a peak value of $\approx 8\%$, and has a full width of half maximum (FWHM) of about 1080\AA\ (Wang et al. 2015a).

Wang et al. (2015a) has minutely described the photometric calibration of LUT for its first 6 months operation,
including the measurements of the throughput at different wavelength in laboratory, the selection of the 
used standard stars from the IUE spectral atlas, the construction of spectral
data sets from NUV to optical band for these standards, the strategy of both observations and data reductions, and the finally reported
magnitude zero point.

The aim of this paper is to investigate the evolution of LUT photometric performance in the
past 18 months by including the new observations taken in one year.
The importance of monitoring the photometric performance of LUT is
described as follows. At first, the magnitude zero point that is used to obtain the actual brightness
of an observed target might change with the time. This change is essential not only for a survey program with a ling time duration, 
but also for
a re-visitation of the same object separated by several months.
Secondly, the study of evolution provides an unique opportunity for the first time to examine the effect on
lunar-based astronomical observations caused by
the extreme environment on the Moon, which is useful for next generation of lunar-based astronomical telescope working in optical/NUV band.

The paper is organized as follows. The adopted calibration strategy is briefly described in
Section 2. The observations and data reductions are given in Section 3. Section 4
presents the results and discussions.

\section{An Outline of Photometric Calibration}

Transforming the observed instrumental magnitude of an
object to its actual brightness requires a pre-determined
magnitude zero point. The zero point (zp) can be determined %by observing of 
through the relationship $\mathrm{zp}=m_{\mathrm{LUT}}-m_{\mathrm{inst}}$ for a series of standard stars,
where $m_{\mathrm{inst}}$ is the observed instrumental magnitude of a standard star and $m_{\mathrm{LUT}}$
its actual brightness. By adopting the AB magnitude system (Oke \& Gunn 1983; Fukugita et al. 1996),
$m_{\mathrm{LUT}}$ is calculated for a given standard as
\begin{equation}
  m_{\mathrm{LUT}}=-2.5\log\frac{\int f_\nu S_\nu d\ln\nu}{\int S_\nu d\ln\nu}-48.6
\end{equation}
where $f_\nu$ is the specific flux density of the standard star in unit of
$\mathrm{erg\ s^{-1}\ cm^{-2}\ Hz^{-1}}$ , and $S_\nu$ the LUT's total throughput at frequency $\nu$ which is
needed to be determined in laboratory at pre-launch (see Figure 1).

\begin{figure}[t]
%\imagei 
\plotone{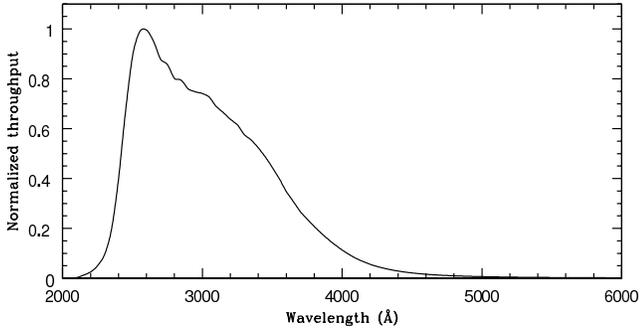}
\caption{The normalized throughput of LUT plotted as a function of wavelength.} %% no full stop at the end
\end{figure}

The IUE standard stars are adopted in our calibration both because of their uniform distribution on sky and because of the limit 
available sky region of LUT. 
Basing upon a combination of the location (i.e., 13.31\symbol{23}S and 14.12\symbol{23}E on the Moon) where it was landed and 
the available range of the gimbal rotation, LUT can only cover a cap around the north pole of the Moon (see Figure 3 in
Wang et al. 2015a). The total available sky area is only about 3600 degrees$^2$, and the 
available sky area at a given time is $\sim400$degrees$^2$.
In total, 44 IUE standards (Wu et al., 1998) located in the available sky region
have been selected by Wang et al. (2015a). 
For each of the standards, its absolute specific flux in NUV band obtained by
IUE has to be extended to optical band (i.e., has a final wavelength coverage from 2000\AA\ to 8000\AA),
both because of the relatively broad wavelength range of LUT (see Figure 1) and because of the red cutoff at 3200\AA\ of the IUE data.
We refer the readers to Section 4.2 in Wang et al. (2015a) for the details of the construction of 
the spectral data sets. Briefly speaking, given the effective temperature, surface gravity and abundance,
the extension is realized by a 3-dimensional linear interpolation basing upon
the ATLAS9 model atmospheres (Castelli \& Kurucz 2003). The absolute specific flux level of each 
extracted model spectrum is determined from its V-band magnitude, after a reddening by its color excess $E(B-V)$.

\section{Observations and Data Reductions}

We have observed 17 standards in 51 observational runs until June 2015.
Table 1 lists the details of the 17 standards.
Columns (7), (8) and (9) tabulates the effective temperature, surface gravity and abundance collected from the literature,
except for two standards whose values are adopted from the suggested models with a solar metallicity for specific stellar types.
Columns (10) lists the magnitudes calculated from Equation (1).
The observations strictly follow the strategy described in Wang et al. (2015a) and Meng et al. (2015), which is designed to
properly remove the strong stray light caused by the sunshine.
Each observational run with a duration of about 30 minutes consists of a series of short exposures with duration of 2 to 10 seconds depending on the
brightness of the standards. The telescope pointing was fixed with
respect to the Moon in the run. This means the shift of a star on the focal
plane within each exposure is negligible compared with the size of the point-spread-function due to the slow rotation of the moon, and 
the total shift within the run is $\sim100$ pixels.

A general pipeline is developed by us to reduce the raw data.
We refer the readers to Meng et al. (2015) for the details of the pipeline.
The procedures of the pipeline includes overscan correction, removal of stray light (including bias and dark current), 
normalization by a composed flat filed, source extraction and cosmicrays rejection
and final aperture photometry. The used composed flat field is derived from a combination of 
the internal flat field provided by the internal LEDs and superflat from a dithering observation of a single standard.
The instrumental magnitude measured with
a fixed radius of 7FWHM is adopted in subsequent analysis, since this aperture
is believed to enclose almost all the signal from a bright star (see Figure 6 in Meng et al. 2015 for a growth curve of
aperture photometry).

\begin{table*}
\small
\caption{IUE standards observed by LUT until June 2015\label{tbl-1}}
\begin{tabular}{@{}lcccccccccc@{}}
\tableline
Star & s.p. type  & $\alpha$(J2000) & $\delta$(J2000) & $m_v$ & $E(B-V)$  & $\mathrm{T_{\mathrm{eff}}}$ & $\log g$ & Fe/H  & $m_{\mathrm{ck04}}$  & Ref. \\
     &            &                 &                 &  mag  &   mag     &    K               &          &             &        mag              &   \\
(1) & (2) & (3) &  (4) & (5) & (6) & (7) & (8) & (9) & (10) & (11) \\
\tableline
HD123299   &   A0III    & 14 04 23.6  &  +64 22 31  &  3.66 & 0.00 & 10371 & 3.95 & -0.19 & 4.61 & 3 \\
HD127700   &   K4IIIBa0.3 & 14 27 31.2 &  +75 41 43  &  4.27 & 0.00 & 4395 &  1.86 & -0.08  & 8.30  & 6\\
HD131873   &   K4III    & 14 50 42.3  &  +74 09 20 &  2.07 & 0.04 & 4077 & 1.7 & -0.10 &  6.52 & 1\\
HD132813   &   M5III    & 14 57 35.6  &  +65 55 54 &  4.59 & 0.00 & 3500 &  1.34 & 0.00 & 9.11 &  \dotfill\\
HD137759   &   K2III    & 15 24 55.7  &  +58 57 57   &  3.29 & 0.01 & 4520 & 2.61 & 0.12 & 7.19 & 3\\
HD139669   &   K5III    & 15 31 25.4  &  +77 20 56  &  4.96 & 0.07 & 3962 & 1.44 & 0.18 & 9.73  & 3 \\
HD147394   &   B5IV     & 16 19 44.4  &  +46 18 45  &  3.90 & 0.01 & 14906 & 4.06 &  0.14 & 3.98  & 1\\
HD153751   &   G5III    & 16 45 58.2  &  +82 02 14 &  4.23 & 0.00 & 5150 & 2.54 & 0.00 &  7.35  &  2 \\
HD159181   &   G2Ib-IIa & 17 30 26.0  &  +52 18 05 &  2.80 & 0.09 & 5325 &  1.51 & -0.02  & 6.09  & 1\\
HD164058   &   K5III    & 17 56 36.4  &  +51 29 20 &  2.23 & 0.01 & 3990 & 1.64 &  0.11  & 6.87   & 3\\
HD166205   &   A1Vn     & 17 32 12.1  &  +86 35 08  &  4.36 & 0.00 & 9230 & 4.10 & 0.00 & 5.63  & \dotfill\\
HD185395   &   F4V      & 19 36 26.5  &  +50 13 16 &  4.48 & 0.00 & 6700 & 4.30  &  0.01  & 6.34  & 4\\
HD188209   &   O9.5Iab  & 19 51 58.9   &  +47 01 39  &  5.63 & 0.20 & 31910 &  3.36 &  0.00  & 5.19  & 5\\
HD188665   &   B5V      & 19 53 17.4  &  +57 31 25 &  5.14 & 0.02 & 14893 & 3.86 & -0.17  & 5.26  & 5  \\
HD198149   &   K0IV     & 20 45 16.6   &  +61 49 38 &  3.42 & 0.01 & 4888 &  3.19  & -0.21 & 6.64   & 3\\
HD214470   &   F3III-IV & 22 35 46.1  &  +73 38 35 &  5.08 & 0.00 & 6637 & 3.59 & 0.09 & 7.18  &  2\\
HD203280   &   A7IV-V   & 21 18 33.6   &  +62 35 05 &  2.46 & 0.00 & 7773 &  3.45 & 0.09 & 4.23   & 7\\
\tableline
\end{tabular}
%% Any table notes must follow the \end{tabular} command.
%\tablenotetext{a}{Sample footnote for table~\ref{tbl-2} that was
%generated with the \LaTeX\ table environment}
%\tablenotetext{b}{Yet another sample footnote for table~\ref{tbl-2}}
%\tablenotetext{c}{Another sample footnote for table~\ref{tbl-2}}
\tablecomments{Column (1): Star name; Column (2): Spectral type; Column (3) \& (4): Right ascension and declination in J2000 mean equator coordinate; 
Column (5): apparent magnitude in V-band; Column (6): Color excess; Column (7): Surface effective temperature; 
Column (8): Surface gravity; Column (9): metal abundance; Column (10): LUT magnitude calculated through Equation (1).
Reference: [1] Koleva \& Vazdekis (2012); [2] Soubiran et al. (2010); [3] Prugniel et al. (2011); [4] Cunha et al. (2000);
[5] Fitzpatrick \& Massa (2005); [6] Luck \& Heiter (2007); [7] Gray et al. (2003).}
\end{table*}

\section{Results and Discussions}

\subsection{Stable Photometric Performance}
Figure 2 shows the variation of the determined magnitude zero point (zp) with time from the beginning of 2014
to June 2015. The value of zp and its uncertainty
in each month are obtained from a statistic on all the observations carried out in the month. All
the errors correspond to a 1$\sigma$ significance level.
One can learn from the
figure that the photometric performance of LUT is highly stable during its 18-months operation.
Based on all the observations of standard until June 2015,
a statistics returns an average value of $\mathrm{zp}=17.53\pm0.05$, which is highly consistent with the results
based on the first 6-months operation (Wang et al. 2015a). The highly stable photometric performance is
further illustrated in Figure 3, which shows the stable light curves (transformed to zp) of four standards that have been repeatedly visited for at least
four times in the past 18 months. 
%The stability of LUT photometric performance can be clearly identified from each light curve.
In addition, one can learn from the figure that the zp values based on different standards are consistent with 
each other within their uncertainties, although a slightly lower value is yielded from the late type-K star HD\,164058 (see discussion given below).

\begin{figure}[t]
%\imagei 
\plotone{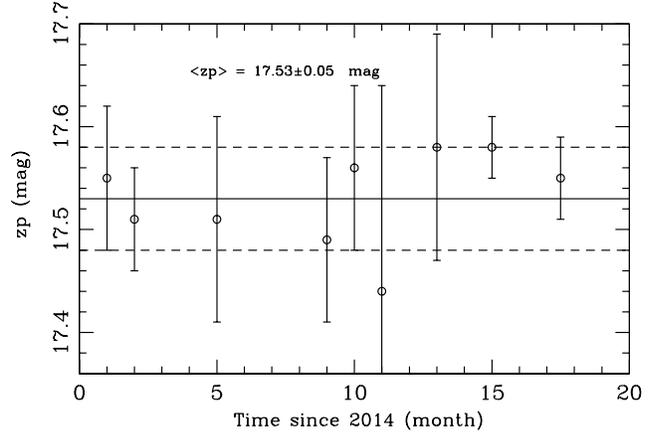}
\caption{The determined magnitude zero point in each month plotted against time in unit of month since the beginning of 2014. The average value of
magnitude zero point is marked by the solid horizontal line. The two dashed horizontal lines mark the corresponding error
at 1$\sigma$ confidence level.} %% no full stop at the end
\end{figure}

\begin{figure}[t]
%\imagei 
\plotone{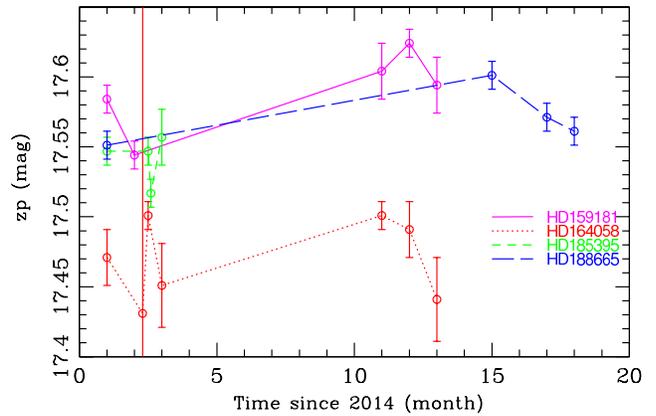}
\caption{The light curves (transformed to magnitude zero point) of four standards that have been repeatedly observed for at least four times in the past 18 months.} %% no full stop at the end
\end{figure}

The revealed highly stable photometric performance therefore indicates that the performance degradation due to the
reaction with the environment is negligible for LUT. 
Possible effects include the bombardment of the high energy protons from solar wind, the oxidation of the coating, and 
the deposition of the charged lunar dust on the the mirrors.

At first, although the damage to the coating on the
mirrors/lens by the protons from solar wind has been alleviated significantly by a designed shielding.
the first reflection mirror of LUT is nevertheless fully exposed to the environment when an
observation is carried out. Assuming a typical inclination of $\theta=$45\symbol{23} with respect to the local zenith, the total
number of protons bombard on the mirror is $N\sim\cos\theta jA\Delta t\sim10^{17}\ \mathrm{p^+}$ in the past 18 months,
where $j\sim10^8\ \mathrm{p^+\ cm^{-2}\ s^{-1}}$ is the proton flux of the solar wind, $A=20\times16.6=332\ \mathrm{cm^{2}}$
the area of the mirror, and $\Delta t$ the total observational time elapsed in the past 18 months.
Secondly, not as the telescopes on the ground or on a low orbit around earth, the degradation of reflection efficiency
due to the oxidation is negligible on the Moon thanks to the extremely tenuous lunar atmosphere (e.g., Stern, 1999; Wang et al. 2011, 2015b).
Finally, the stable  photometric performance in the past 18 months suggests that LUT did not suffer the influence of the charged lunar dust that can
continuously decrease the reflection/transparency efficiency by a deposition on the mirrors, although this effect has been indeed observed by
the Apollo 12 astronaut (e.g., Colwell et al. 2007). The avoidance of the influence of the dust
can be understood by the following two reasons. At the beginning, LUT is mounted within a cabin of
the CE-3 lander. Secondly, for each lunar day, the cabin was closed at both sunrise and sunset when the dust particles are believed to
be launched at terminator (e.g., Berg et al. 1976).

\subsection{Uncertainties}

The calibration based on the IUE standards results in a magnitude zero point with an accuracy of 5\%.  
The resulted final uncertainty contains the contributions from the shot noise from incident light, the residual nonuniformity in 
CCD quantum efficiency after the flatfielding, the error in the determined throughput curve,
and the modeled standard spectra (i.e., the uncertainties of the parameters defining individual stellar atmosphere, see discussion given in Section 4.3). 
Because the used standards are so bright, the shot noise is generally estimated to be as low as 0.001-0.003 magnitudes for these observed 
standards. The magnitude uncertainty caused by the error of the throughput curve is estimated from a Monte-Carlo simulation with 100 iterations,
in which a random curve is produced by a random sampling at each wavelength according to the measured error of $S_\lambda$.
The simulation shows that the resulted magnitude error is $\sim0.001$ mag.
% which can be easily understood because the magnitude is 
%obtained from an integration of the curve. 
As indicated by Meng et al. (2015), the residual medium-scale nonuniformity in CCD contributes 
a typical error of $\sim0.02$mag for a 10 mag star. 
The finally used flat field comes from a combination of the intrinsic flat field
and the super flat field. That means only the nonuniformity in CCD quantum efficiency on the scale of pixels and scale of hundreds of pixels
can be corrected by the intrinsic flat field and the super flat field, respectively.

\subsection{Independence On Stellar Spectral Type}
To investigate the dependence of the determined zp value on the spectral type of the standard from which zp is determined,
we separate the observed 17 standards into three groups according to their spectral types. 
Table 2 lists the zp values that are determined from each group, which allows us to draw a conclusion that the photometric calibration of LUT is 
independent on spectral type of the used standard.  
% which shows that the zp values determined from the
%standards with different spectral types are consistent with each other within the uncertainties.
The table additionally shows a trend in which the estimated uncertainty of zp increases with spectral type of the used standard:
later the spectral type, larger the resulted uncertainty will be. This trend is not hard to be understood.
Except the imperfect flatfielding in a middle scale of tens of pixels discussed above, 
the error of the determined zp is additionally contributed by the uncertainty of the calculated magnitude of a standard.
In addition to depend on the IUE spectral/ATLAS9 model calibration, the calculated magnitude depends on the V-band magnitude, the level of extinction and
the parameters defining the atmosphere model. The level of this dependence is
expected to be more significant for a late type star than for an early type star, because the peak of the
SED of a star shifts toward a longer wavelength when the stellar surface temperature decreases.

\begin{table}
\small
\caption{The resulted magnitude zero points from the standards with different spectral types.\label{tbl-2}}
\begin{tabular}{@{}ccc@{}}
\tableline
s.p. type  &  zp  & Num. of runs\\
     &   mag     &     \\
(1) & (2)  & (3)\\
\tableline
O, B, \& A & $17.55\pm0.04$ & 11\\
F \& G     & $17.60\pm0.07$ & 17\\
K \& M     & $17.45\pm0.13$ & 23\\
\tableline
\end{tabular}
\tablecomments{Column (1): Spectral type; Column (2): Determined \\
magnitude zero point; Column (3): Total number of the used \\
observational runs.}
%% Any table notes must follow the \end{tabular} command.
%\tablenotetext{a}{Sample footnote for table~\ref{tbl-2} that was
%generated with the \LaTeX\ table environment}
%\tablenotetext{b}{Yet another sample footnote for table~\ref{tbl-2}}
%\tablenotetext{c}{Another sample footnote for table~\ref{tbl-2}}
\end{table}

\section{Summary}

The photometric performance of LUT is reported here for
its 18-months operation on the Moon. The observations of 17 IUE standards in 51 runs allow us to
claim a highly stable photometric performance of LUT during the past 18 months.
The magnitude zero point is determined to be $17.53\pm0.05$ mag, and
is found to be independent on the spectral type of the used standard.

%\paragraph{A short head four}
%Ui posuit fines tuos pacem,
%et adipe frumenti satiat te. Qui emittit eloquium suum terrae;
%velociter currit sermo eius.  Qui dat nivem sicut lanam, nebulam
%sicut cinerem spargit. Mittit crystallum suam sicut buccellas; ante
%faciem frigoris eius quis sustinebit. Emittet verbum suum, et
%liquefaciet ea; flabit spiritus eius, et fluent aquae.

%\subparagraph{A short head five leads into paragraph.} Ui posuit fines tuos pacem,
%et adipe frumenti satiat te. Qui emittit eloquium suum terrae;
%velociter currit sermo eius.  Qui dat nivem sicut lanam, nebulam
%sicut cinerem spargit. Mittit crystallum suam sicut buccellas; ante
%faciem frigoris eius quis sustinebit. Emittet verbum suum, et
%liquefaciet ea; flabit spiritus eius, et fluent aquae.

\acknowledgments
We would like to thank the anonymous referee for his/her suggestions that improve the manuscript.
The authors thank the outstanding work of the LUT team and support by the
team from the ground system of the Chang’e-3 mission. This study is supported by the Key Research
Program of Chinese Academy of Sciences (KGED-EW-603). JW is supported by the National
Natural Science Foundation of China under Grant 11473036. MXM is supported by the National
Natural Science Foundation of China under Grant 11203033.

%Use \verb!\cite! command to cite reference(s).
%\smallskip

%\noindent
%\verb!\cite{bag02}! -- \cite{bag02}\\
%\verb!\citep{bag02}! -- \citep{bag02}\\
%\verb!\citet{bag02}! -- \citet{bag02}\\
%\verb!\cite{ale94}! -- \cite{ale94}\\
%\verb!\cite{bag02,ale94}! -- \cite{bag02,ale94}\\
%\verb!\citeauthor{ale94}! -- \citeauthor{ale94}\\
%\verb!\citeyear{ale94}! -- \citeyear{ale94}\\

\nocite{*}
%\bibliographystyle{spr-mp-nameyear-cnd}
%\bibliography{myref}
%\bibliography{biblio-u1}

\begin{thebibliography}{}

\bibitem[Berg et al. 1976]{ber76} Berg, O. E., Wolf, H., \& Rhee, J.: Lunar and Planetary Exploration 48, 233 (1976)
\bibitem[Cao et al. 2011]{cao11} Cao, L., Ruan, P., Cai, H.B., et al.: ScChG 54, 558 (2011)
\bibitem[Castelli \& Kurucz 2003]{cak03} Castelli, F. \& Kurucz, R.: IAU Symposium 210, Modelling of  Stellar Atmospheres, Uppsala, Sweden, eds. N.E. Piskunov,
W.W. Weiss. and D.F. Gray, ASP-S210 (2003)
\bibitem[Colwell et al. 2007]{col07} Colwell, J. E., Batiste, S., Horanyi, M., Robertson, S., \& Sture, S.: Reviews of Geophysics 45, 2006 (2007)
\bibitem[Cunha et al. 2000]{cun00} Cunha, K., Smith, V. V., Boesgaard, A. M., \& Lambert, D. L.: \apj\ 530, 939 (2000)
\bibitem[Fitzpatrick \& Massa 2005]{fim05} Fitzpatrick, E. L., \& Massa, D.: \apj\ 129, 1642 (2005)
\bibitem[Fukugita et al. 1996]{fuk} Fukugita, M., Ichikawa, T., Gunn, J.E., et al.: \aj\ 111, 1748 (1996)
\bibitem[Gray et al. 2003]{gra03} Gray, R. O., Corbally, C. J., Garrison, R. F., McFadden, M. T., \& Robinson, P. E.: \aj\ 126, 2048 (2003)
\bibitem[Ip et al. 2014]{ip14} Ip, W.-H., Yan, J., Li, C.-L., Ouyang, Z.-Y.: Research in Astronomy and Astrophysics 14, 1511 (2014)
\bibitem[Koleva \& Vazdekis 2012]{kov12} Koleva, M., \& Vazdekis, A.: \aap\ 538, 143 (2012)
\bibitem[Luck \& Heiter 2007]{luc07} Luck, R. E., \& Heiter, U.: \aj\ 133, 2464 (2007)
\bibitem[Meng et al. 2015]{men15} Meng, X. M., et al.: astro-ph/arXiv:1505.07951 (2015)
%\bibitem[Martin et al. 2005]{mar05} Martin, D. C., et al. 2005, ApJL, 619, 1
%\bibitem[Morrissey et al. 2005]{mor05} Morrissey, P., Schiminovich, D., Barlow, T.A., et al. 2005, ApJL, 619, 7
%\bibitem[Nichols \& Linsky 1996]{nil96} Nichols, J. S., \& Linsky, J. L. 1996, AJ, 111, 517
\bibitem[Oke \& Gunn 1983]{okg83} Oke, J. B., \& Guun, J. E.: \apj\ 266, 713 (1983)
\bibitem[Prugniel et al. 2011]{pru11} Prugniel, P., Vauglin, I., \& Koleva M.: \aap\ 531, 165 (2011)
\bibitem[Schneider et al. 1983]{sch83} Schneider, D. P., Gunn, J. E., \& , Hoessel, J. G.: \apj\ 264, 337 (1983)
\bibitem[Soubiran et al. 2010]{sou10} Soubiran, C., et al.: \aap\ 515, 111 (2010)
\bibitem[Wang et al. 2011]{wan11} Wang, J. et al.: AdSpR 48, 1927 (2011)
\bibitem[Wang et al. 2015a]{wan15a} Wang, J. et al.: Research in Astronomy and Astrophysics 7, 1068 (2015a)
\bibitem[Wang et al. 2015b]{wan15b} Wang, J. et al.: P\&SS 109, 123 (2015b)
\bibitem[Wu et al. 1998]{wu98} Wu, C. C., et al.: ESASP 413, 751 (1998)
\end{thebibliography}

\end{document}